\documentclass{article}
\usepackage{amsmath}
\usepackage[version=4]{mhchem}

\usepackage[utf8]{inputenc} 
\usepackage[T1]{fontenc}    
\usepackage{hyperref}       
\usepackage{url}            
\usepackage{booktabs}       
\usepackage{amsfonts}       
\usepackage{nicefrac}       
\usepackage{microtype}      
\usepackage{xcolor}         

\usepackage{graphicx}
\usepackage{authblk}

\title{Model Order Reduction for Quantum Molecular Dynamics}

\author[1]{Siu Wun Cheung}
\author[1]{Youngsoo Choi}
\author[2]{Jean-Luc Fattebert}
\author[1]{Daniel Osei-Kuffuor}

\affil[1]{Center for Applied Scientific Computing, Lawrence Livermore National Laboratory, Livermore, CA 94550}
\affil[2]{Computational Sciences \& Engineering Division, Oak Ridge National Laboratory, Oak Ridge, TN 37830}

\begin{document}

\maketitle

\begin{abstract}
Quantum molecular dynamics (QMD) provides a high-fidelity physical framework for modeling atomic and molecular phenomena from first principles. However, its mathematical formulation requires solving iterative, high-dimensional electronic structure problems at every physical state step. Within the broader data-driven computational mechanics paradigm, we present a projection-based model order reduction framework for electronic structure calculation in QMD. By discovering a low-dimensional manifold of the electronic solution space within Kohn-Sham density functional theory, our approach projects the governing variational formulation onto an intrinsically reduced subspace. This avoids full-space iterative wavefunction optimization while preserving rigorous physical symmetries and state consistency. We demonstrate this methodology on a water molecule, demonstrating strong agreement with high-fidelity reference solutions across structural trajectories and thermodynamic observables through canonical coordinate alignment and structured manifold sampling.
\end{abstract}

\section{Introduction}

Molecular dynamics (MD)~\cite{frenkel2023understanding} is a fundamental computational mechanics tool for probing the spatio-temporal evolution of material systems, chemical processes, and complex physical media. Within the data-driven modeling community, significant attention has centered on machine learning interatomic potentials (MLIPs)~\cite{deringer2019machine}. These statistical approaches learn explicit maps from atomic coordinates to energy surfaces, effectively substituting continuum or quantum mechanics solves with surrogate evaluations. While successful for standard dynamics, MLIPs often struggle with out-of-distribution transferability because the calculated forces do not originate from the explicit minimization of a physical energy functional at each step. This can limit their structural consistency under novel mechanical or thermodynamic conditions.

Quantum molecular dynamics (QMD)~\cite{marx2009ab} offers a fundamental alternative by solving the coupled nucleo-electronic problem directly from quantum mechanical principles. Its primary computational structure requires solving a nonlinear electronic structure eigenvalue problem, typically governed by Kohn-Sham (KS) Density Functional Theory (DFT)~\cite{cances2023density}, at every time step. The continuous high-dimensional discretization of these electronic states forms the primary computational challenge in standard QMD implementations.

In this work, we propose a projection-based model order reduction (MOR) paradigm for electronic structure mechanics. Drawing from physics-preserving projection methods established in solid, fluid, and multiphysics mechanics~\cite{choi2021space, brunini2022projection, copeland2022reduced, cheung2023local, tsai2023accelerating, chung2024scalable, rathore2025projection, youkilis2025projection}, our framework projects the governing KS DFT operator onto a low-dimensional reduced subspace constructed from snapshot data. This maintains first-principles physical fidelity while reformulating the high-dimensional electronic state solve into a reduced subspace optimization. By combining systematic parameter space sampling with symmetry-aware canonical reference frame transformations, we demonstrate that data-driven reduced spaces can capture complex trajectory dynamics while strictly obeying fundamental physical conservation laws.

\section{Quantum Molecular Dynamics Mechanics}
\label{sec:qmd}

QMD governs the coupled temporal evolution of atomic nuclei and electronic fields. Under the Born-Oppenheimer separation, the lighter electronic field rapidly relaxes to its ground state for any instantaneous nuclear configuration $\{\mathbf{R}_I\}_{I=1}^{N_I}$. The nuclear trajectories are updated using numerical time-stepping schemes such as the velocity Verlet algorithm:
\begin{equation}
\mathbf{R}_I(t + \Delta t) = 2\mathbf{R}_I(t) - \mathbf{R}_I(t - \Delta t) + \frac{\Delta t^2}{M_I} \mathbf{F}_I(t),
\end{equation}
where $M_I$ denotes the mass of nucleus $I$, and $\mathbf{F}_I(t)$ represents the force field driving structural motion. Forces are derived from the variational gradient of the total electronic energy functional with respect to nuclear positions:
\begin{equation}
\mathbf{F}_I = -\nabla_{\mathbf{R}_I} E[ \{ \phi_i \}_{i=1}^N ],
\end{equation}
where $E[\{ \phi_i \}_{i=1}^N]$ depends explicitly on the ground state electronic wavefunctions $\{ \phi_i \}_{i=1}^N$.

In KS DFT, these wavefunctions satisfy a coupled system of single-particle equations:
\begin{equation}
H \phi_i = \epsilon_i \phi_i,
\label{eq:ks_brief}
\end{equation}
where $H$ is the nonlinear KS Hamiltonian operator, $\phi_i$ are the spatial orbitals, and $\epsilon_i$ are orbital energies. The Hamiltonian depends continuously on nuclear positions $\{\mathbf{R}_I\}_{I=1}^{N_I}$ and the spatial electron density $\rho_e(\mathbf{r}) = 2 \sum_{i=1}^{N} f_i \vert \phi_i(\mathbf{r}) \vert^2$, with state occupations $f_i \in [0,1]$. This system creates a nested mechanics structure: time integration on the outside, self-consistent field (SCF) fixed-point iterations in the middle, and continuous spatial discretization solvers at the core.

\section{Data-Driven Subspace Reduction}
\label{sec:mor}

To mitigate the computational cost of continuous spatial discretizations, we formulate a data-driven projection approach that models the electronic state solution manifold within a compact basis.

A key aspect of this reduction is parameter space alignment. For molecular assemblies, the electronic field exhibits SE(3) equivariance with respect to rigid body translations and rotations. We map arbitrary spatial states into a fixed canonical reference frame. For a water molecule, anchoring the oxygen atom at the coordinate origin and orienting its principal axes isolates the intrinsic kinematic degrees of freedom, specifically bond lengths and bond angles. This parameterization compresses the snapshot domain, enabling full coverage of the active manifold with limited sampling points.

\begin{figure}[h!]
    \centering
    \includegraphics[width=0.48\linewidth]{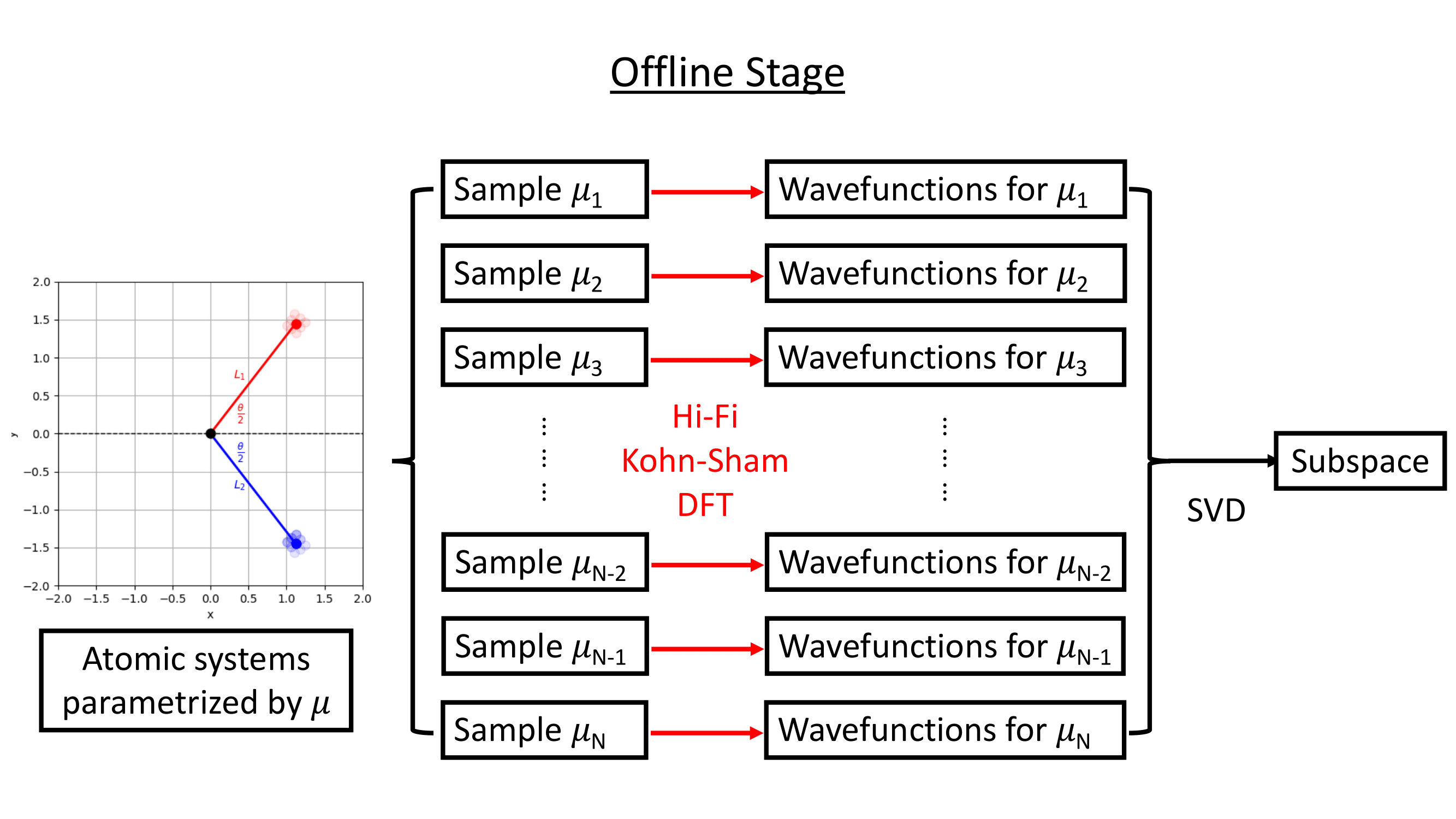}
    \hspace{0.02\linewidth}
    \includegraphics[width=0.48\linewidth]{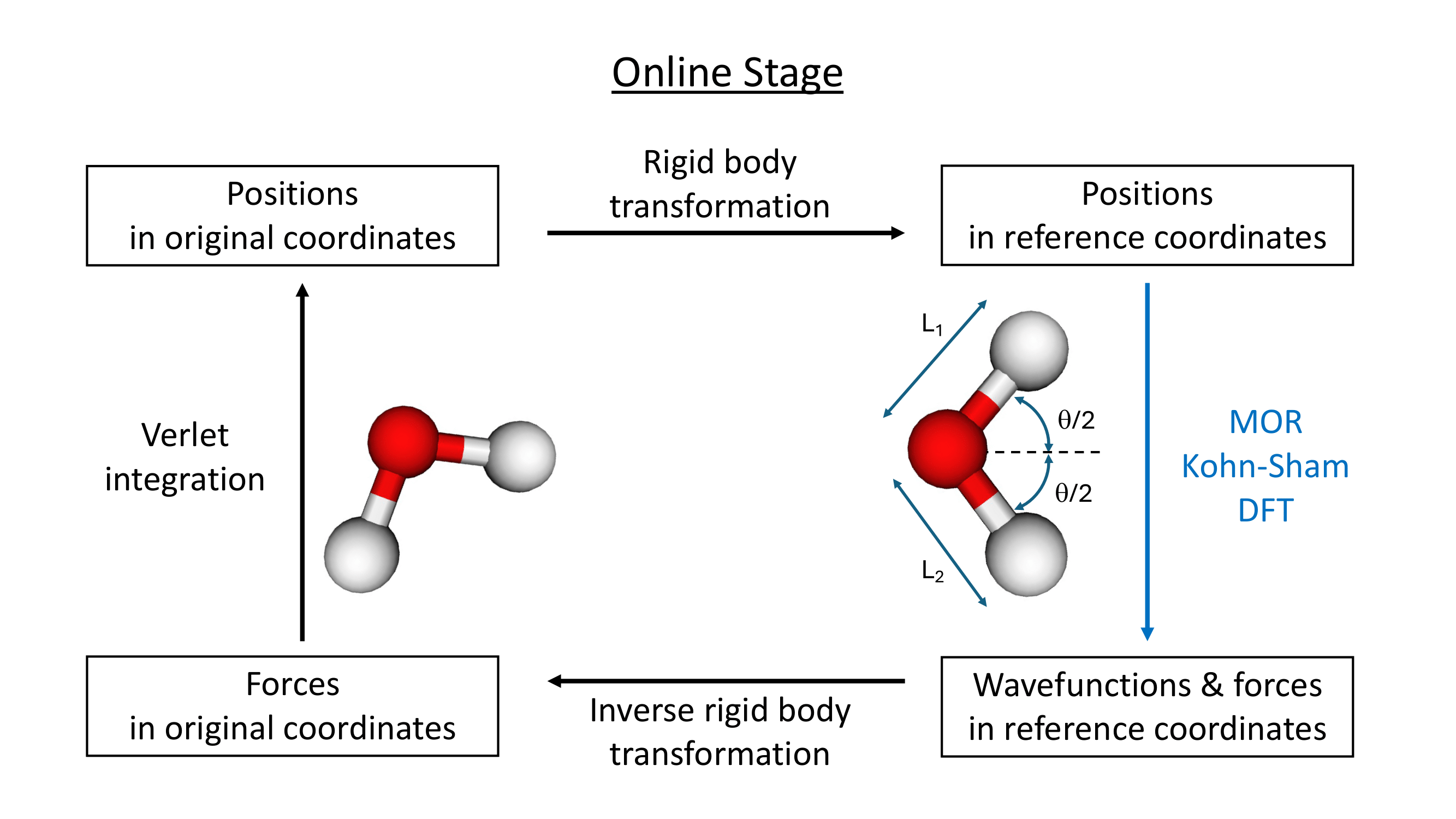}
    \caption{Data-driven reduced order modeling workflow: offline canonical snapshot generation and online projection-based dynamic trajectory integration.}
    \label{fig:schematic}
\end{figure}

The projection workflow separates into offline basis generation and online reduced solves:

\textbf{Offline Phase:} We sample the canonical configuration space as depicted in Figure~\ref{fig:schematic}. High-fidelity (Hi-Fi) electronic structures are solved using an accelerated block preconditioned gradient solver (ABPG)~\cite{FATTEBERT2010abpg} to collect snapshot matrices. Proper Orthogonal Decomposition via snapshot Singular Value Decomposition (SVD) extracts a low-dimensional orthonormal basis spanning the electronic solution subspace, truncated according to energy fraction criteria.

\textbf{Online Phase:} During physical trajectory integration, unaligned atomic configurations are mapped to the canonical reference frame via rigid body transformations (Figure~\ref{fig:schematic}). The electronic structure state is solved via Mermin free energy minimization projected directly onto the pre-computed subspace. State occupation numbers are updated via ensemble minimization over frozen wavefunctions~\cite{marzari1997ensemble}, requiring only reduced-dimension eigenvalue solves~\cite{cheung2024theory}. Resulting forces are computed within the reference frame and transformed back to physical space to integrate nuclear positions via Verlet steps.

\section{Numerical Experiments}
\label{sec:experiments}

We evaluate the fidelity of this subspace formulation on a standard water molecule (\ce{H2O}) system containing 3 nuclei and 8 valence/core electronic states, mapped to $N=4$ doubly occupied spatial orbitals. The system is defined in a canonical frame with the oxygen atom fixed at the origin, one hydrogen atom \ce{H1} positioned in the first quadrant, and the second hydrogen \ce{H2} located in the fourth quadrant.

The system geometry is parameterized by two bond lengths ($L_1 = 1.83s_1$ Bohrs for $s_1 \in [0.95, 1.05]$, and $L_2 = 1.83s_2$ Bohrs for $s_2 \in [0.95, s_1]$) and one bond angle ($\theta = 104.5 + s_\theta$ degrees for $s_\theta \in [-5, 5]$). Uniform grid sampling across this parametric domain yields 18 training configurations, producing 72 discrete orbital snapshots. High-fidelity discretization utilizes a fourth-order finite difference operator on a $64^3$ grid ($262,144$ spatial degrees of freedom). Snapshot SVD yields a 34-dimensional reduced subspace capturing $0.9999$ of the total energy fraction, reducing the continuous spatial state dimension from over $10^5$ down to 34 reduced variables.

\begin{figure}[h!]
    \centering
    \includegraphics[width=0.32\linewidth]{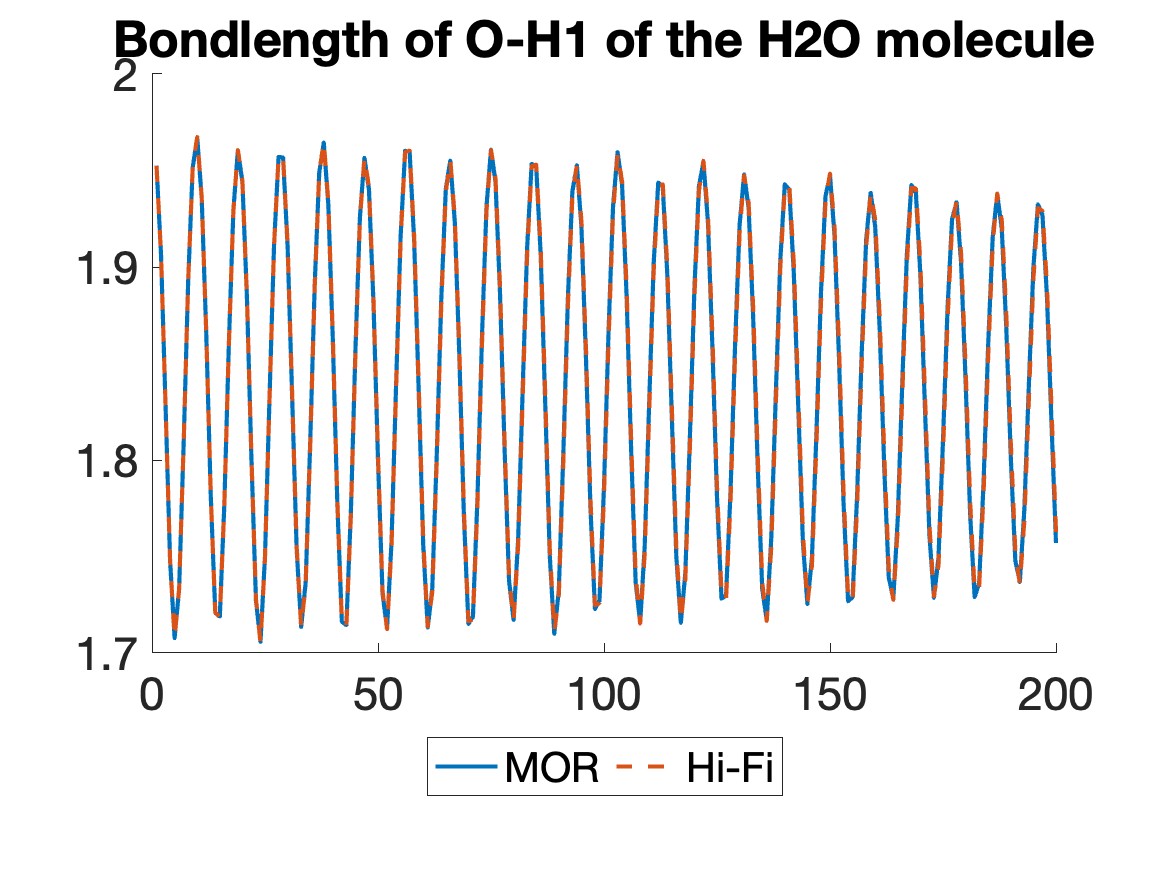}
    \includegraphics[width=0.32\linewidth]{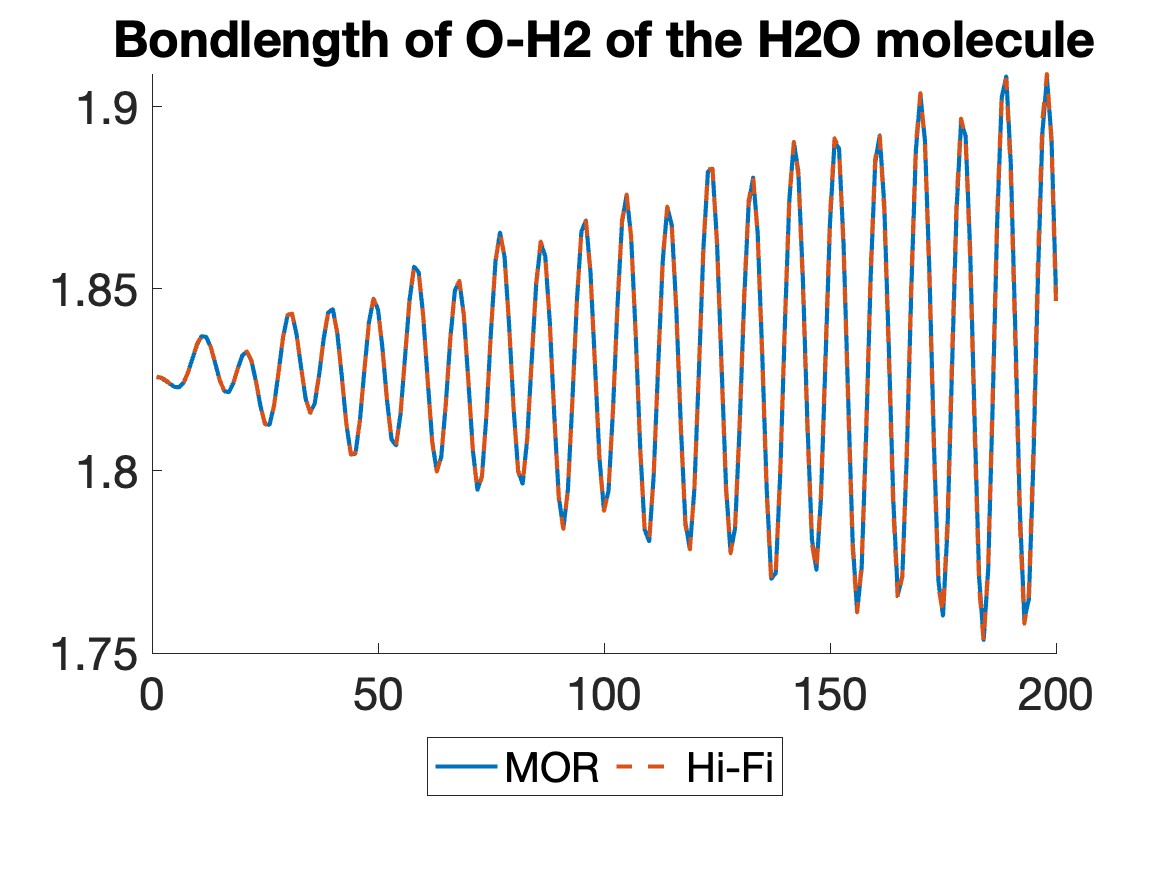}
    \includegraphics[width=0.32\linewidth]{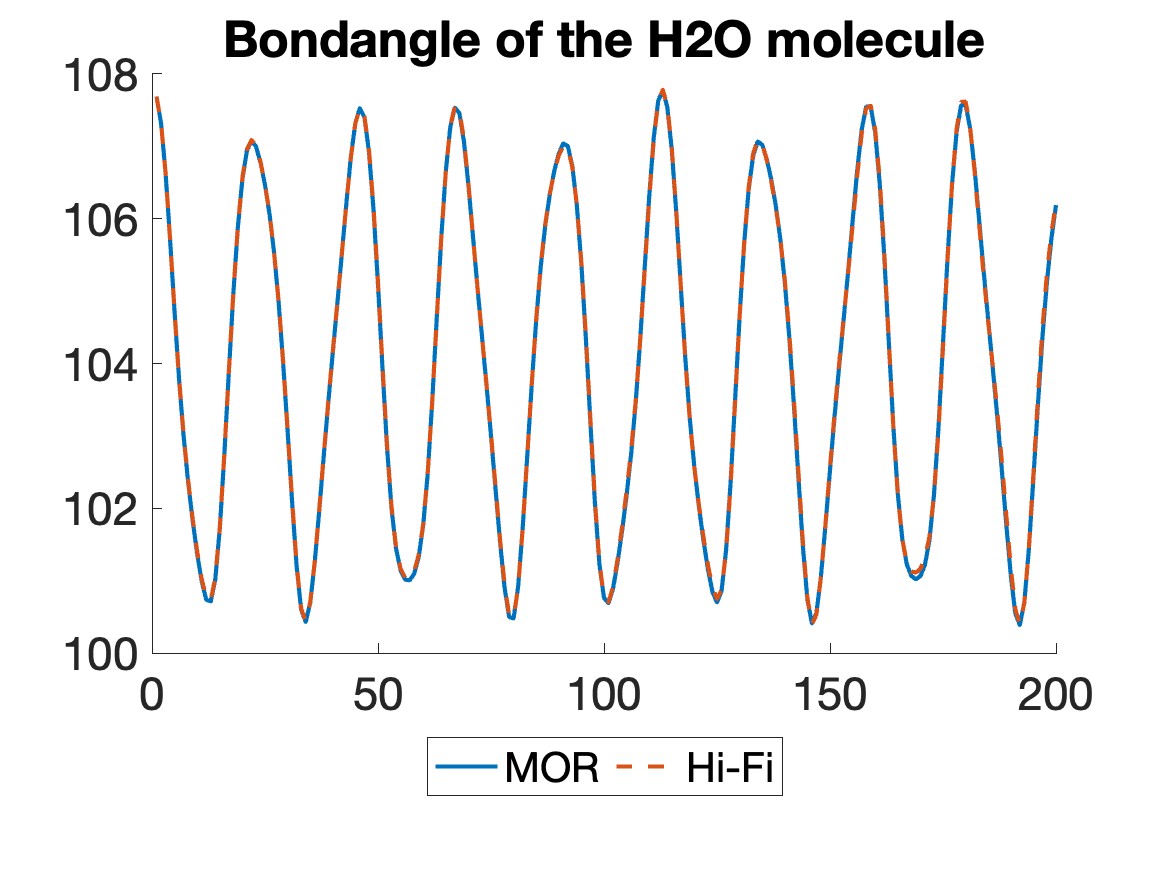}
    \caption{Trajectory comparison of geometric structural parameters between full-space Hi-Fi solutions and the data-driven reduced subspace model.}
    \label{fig:bond_results}
\end{figure}

We test the reduced subspace model on an unconstrained thermal trajectory at $300\,\text{K}$. Figure~\ref{fig:bond_results} tracks the time evolution of the structural parameters: \ce{O-H1} bond length, \ce{O-H2} bond length, and \ce{H1-O-H2} bond angle. The projection-based MOR model (solid blue) closely tracks the high-fidelity QMD reference dynamics (dashed red), accurately capturing internal vibrational modes.

\begin{figure}[h!]
    \centering
    \includegraphics[width=0.32\linewidth]{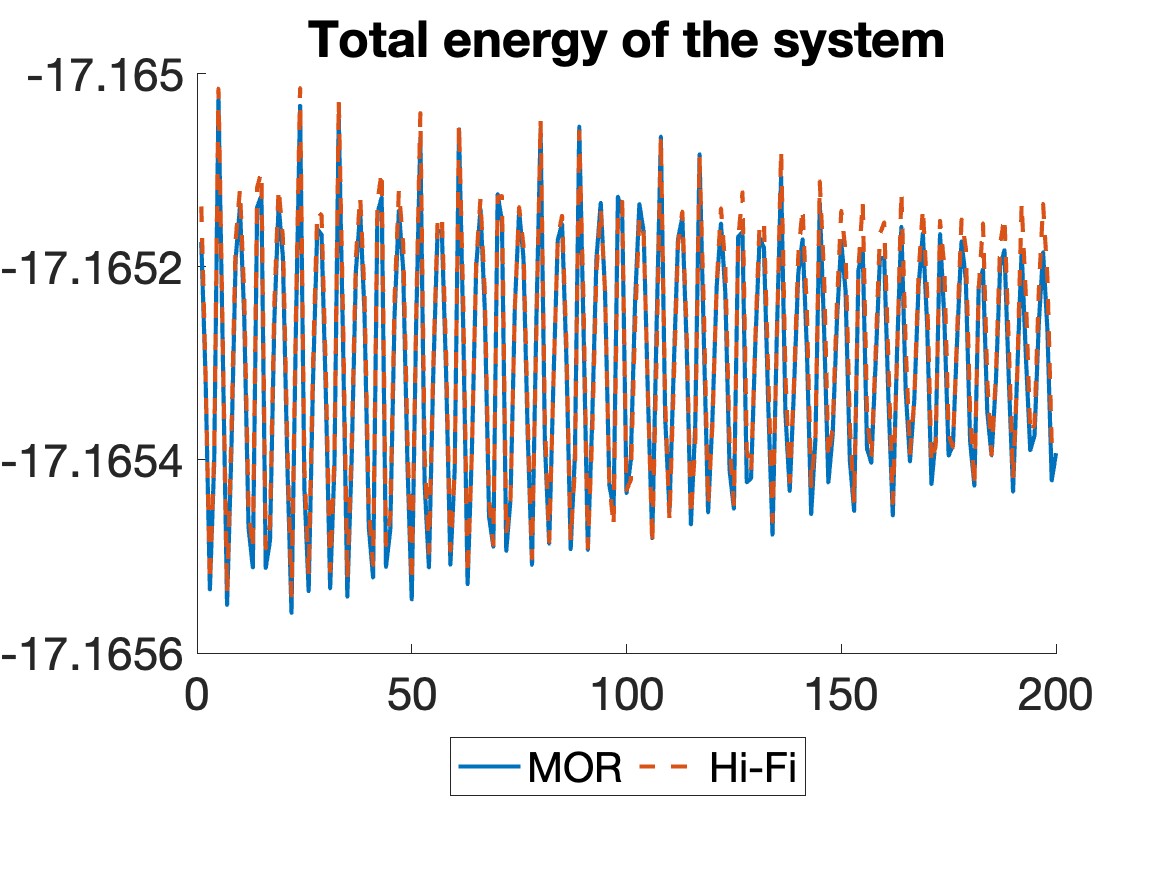}
    \includegraphics[width=0.32\linewidth]{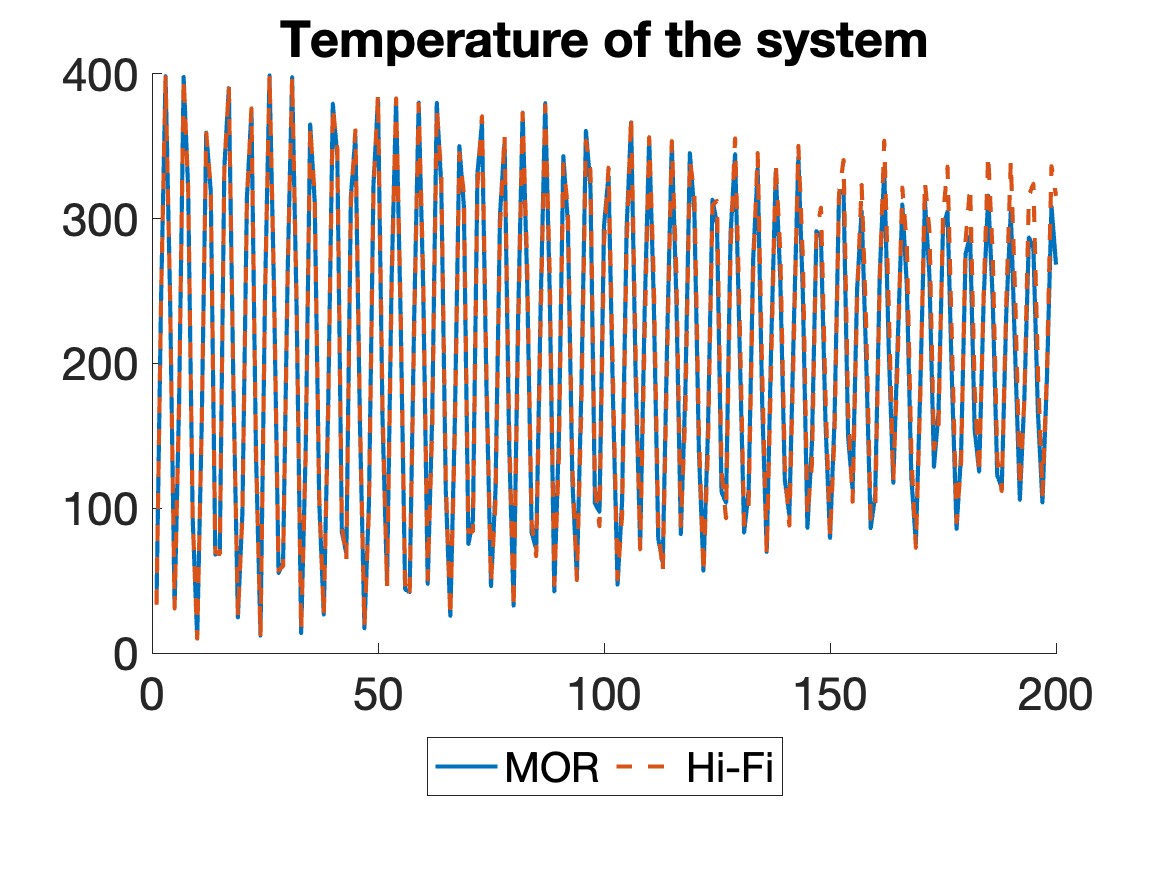}
    \caption{Thermodynamic state variables and physical invariants over dynamic trajectory steps.}
    \label{fig:system_results}
\end{figure}

Figure~\ref{fig:system_results} shows the system thermodynamic properties, including kinetic energy, total energy, and instantaneous temperature. The MOR formulation maintains total energy conservation and accurate thermal fluctuations over time, demonstrating that subspace projection can preserve key physical invariants during dynamic integration.

\section{Conclusions and Open Directions}
\label{sec:conclusion}

This work demonstrates a projection-based model order reduction formulation for electronic structure calculations in quantum molecular dynamics. By combining canonical frame transformations with snapshot manifold discovery, the method reduces continuous spatial wavefunctions to a low-dimensional subspace while maintaining first-principles consistency. Numerical validation confirms that projected reduced solves can reproduce structural mechanics and thermodynamic states without requiring full-space iterative updates at every dynamic step.

These results point toward several open research directions in data-driven computational mechanics. A promising avenue is the development of modular domain decomposition methodologies, such as a "Local Learning, Global Assembly" framework~\cite{choi2025defining}. By constructing localized, transferrable subspace bases on subdomains and assembling them globally across heterogeneous physical structures, reduced order computational mechanics can expand to complex material interfaces, non-equilibrium phase transformations, and multi-scale physical systems.

\section*{Acknowledgement}
This work was supported by Laboratory Directed Research and Development (LDRD) Program by the U.S. Department of Energy (24-ERD-035). 
Lawrence Livermore National Laboratory is operated by Lawrence Livermore National Security, LLC, for the U.S. Department of Energy, National Nuclear Security Administration under Contract DE-AC52-07NA27344. IM release number: LLNL-CONF-2010261.
This manuscript has been co-authored by UT-Battelle, LLC, under contract DE-AC05–00OR22725 with the US Department of Energy (DOE).

\bibliographystyle{unsrt}
\bibliography{references}

\end{document}